\documentclass[floatfix,twocolumn,superscriptaddress,prl,aps]{revtex4}
\usepackage{graphicx}
\begin{document}
\bibliographystyle{apsrev}
{\bf Eggert, Affleck, and Horton Reply to the ``Comment on 'Neel order in doped quasi one-dimensional antiferromagnets' ''\\}
\author{Sebastian Eggert}
\affiliation{Institute of Theoretical Physics,
Chalmers University of Technology and G\"oteborg University,
S-412 96 G\"oteborg, Sweden}
\author{Ian Affleck}
\affiliation{Physics Department, Boston University,
 Boston, MA 02215}
\altaffiliation[On leave from ]{Canadian Institute for 
Advanced Research and Department of Physics
 and Astronomy, University of British Columbia, Vancouver,
BC,  Canada, V6T 1Z1}
\author{Matthew D.P. Horton}
\affiliation{99 John Street, New York, NY 10038}
\date{\today}
In the Comment \cite{comment} it is pointed out correctly that 
the field theory treatment that was used in our recent Letter \cite{letter}
to obtain some of the results for the Heisenberg antiferromagnetic chain
is indeed only valid in the limit 
of long length $L$, low temperature $T$, and small magnetization $S^z$.
In particular, this treatment becomes only asymptotically 
correct in a region where the dispersion is linear and the
spin-wave velocity $v$ can be approximated by a constant \cite{ourPRB}, 
which according to our numerics is the case if both $T \alt 0.2 J$ 
and $L \agt 10$ sites.  There is no restriction on
the product $LT/v$ as long as $v$ is approximately constant.

\begin{figure}
\includegraphics[width=.44\textwidth]{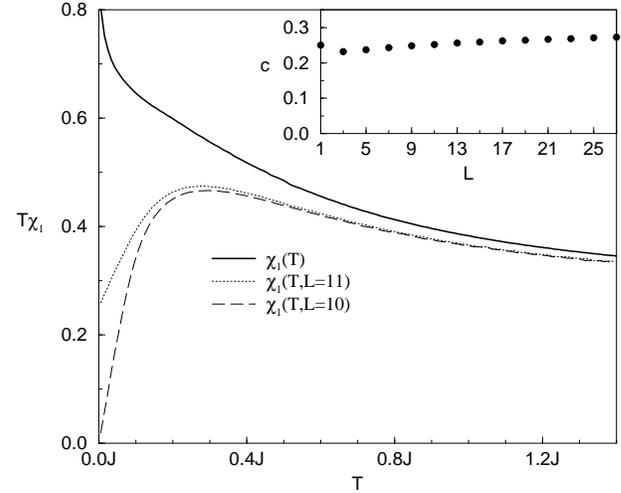}
\caption{The staggered susceptibility $\chi_1(T)$ in the 
thermodynamic limit determined by combining 
bosonization results at lower temperature and 
numerical simulations at higher temperature. The  numerical results for
$L=10$ and $L=11$ are also shown.  Inset:  the intercept 
$c=\lim_{T\to 0}T \chi_1(T,L)$ as a function of $L$.}
\label{chi1-fig}
\end{figure}

However, we must emphasize that we were indeed able to calculate the 
staggered susceptibility $\chi_1$
for {\it arbitrary} $L$ and $T$ as mentioned in
the introduction  by combining
the field theory results with numerical calculations \cite{note}. The  
numerical 
calculations are especially reliable for values of $L$ and $T$ where 
the field theory predictions become invalid and vice versa. 
We can therefore describe the entire crossover of $\chi_1$
to the limit of large $T$ and/or small $L$, which shows an
interesting behavior by itself that was unfortunately not explicitly presented
in the Letter \cite{letter}.  If we for example consider the 
staggered susceptibility $\chi_1$
without impurities as a function of $T$ we see that it crosses 
over from the bosonization formula 
to a high temperature expansion as shown in Fig.~\ref{chi1-fig}.
\begin{equation}
\chi_1(T) {\longrightarrow}\   
\left\{
\begin{array}{lcl}
\frac{b \sqrt{ln(a/T)}}{T} &\phantom{nnn}& T \ll J\\
& & \\
\frac{1+J/2T}{4 T} &\phantom{nnn}& T \gg J\\
\end{array}\right.
\label{chi1}
\end{equation}
where $a \sim 23J$ and $b = \frac{\Gamma^2(1/4)}{4 \sqrt{2 \pi^3}\Gamma^2(3/4)}
\approx 0.277904$.  In  the case of shorter chain lengths 
$L$ we again find a significant drop from the thermodynamic
limit as well as a split at $T \alt 4J/L$ for even and odd chains as depicted
for $L = 10$ and $L = 11$ in Fig.~\ref{chi1-fig}. The crossover from
finite size behavior to the thermodynamic limit is therefore
very similar to Fig.~1 in our Letter \cite{letter} which shows 
the behavior predicted by bosonization in the limit $L\to \infty$, 
$T\to 0$ as a function of $LT$, compared to numerical 
results for large $L$. Even for smaller $L$ we find again that
$\chi_1(T,L) \propto L$ for even chains as $T\to 0$ 
and $\chi_1(T,L) \to c/T$ for odd 
chains, where
the intercept $c$ can be approximated by a length independent constant 
even down to $L=1$ as shown in the inset of Fig.~\ref{chi1-fig}.

Now that we have displayed $\chi_1$  for arbitrary $T$
we may be tempted to again apply the chain mean field equation
\begin{equation}
z J' \chi_1(T_N) = 1
\end{equation}
even in the case where $J'$ is of the order of $J$.  Although we
might not expect any one-dimensional physics to survive in that limit,
we find for example that this would result in $T_N \approx 1.386 J$ for
a simple cubic lattice with $J=J'$, which is indeed higher than the 
accepted values \cite{pan}, but still an improvement over the
ordinary  mean field result of $T_N =1.5J$.  
If $J'$ is of order $J$ only extreme doping levels will
significantly affect the ordering temperature, since finite size
effects are small at higher temperatures $T  \agt 4J/L$. 
In conclusion we have calculated the staggered susceptibility for 
arbitrary $L$ and $T$ and outlined in more detail the behavior 
in the limit of large $T$ and small $L$.\\~\\
Sebastian Eggert, Ian Affleck, and Matthew D.P.~Horton


\begin{thebibliography}{999}
\bibitem{note}  We never claimed that the results for arbitrary $T$ and $L$
were obtained solely by bosonization methods as clearly outlined on the top of page 2 in the Letter.
\end{thebibliography}
\end{document}